\documentclass[prl,aps,twocolumn,groupedaddress,floats,showpacs,final]{revtex4-1}
\usepackage{graphicx}
\usepackage{dcolumn}
\usepackage{bm}
\usepackage{color}
\definecolor{blue}{rgb}{0.3,0.3,0.9}
\def\he4{$^4$He}
\def\beq{\begin{equation}}
\def\eeq{\end{equation}}

\begin{document}

\title{Topological quantum phases  of \he4 confined to nanoporous materials}

\author{Lode Pollet}
\affiliation{Department of Physics, Arnold Sommerfeld Center for Theoretical Physics and Center for NanoScience, University of Munich, Theresienstrasse 37, 80333 Munich, Germany}
\author{Anatoly B. Kuklov}
\affiliation {Department of Engineering Science and Physics, CUNY, Staten Island, NY 10314, USA}

\date{\today}
\begin{abstract}
The ground state of \he4 confined in a system with the topology of cylinder  can display properties of solid, superfluid and liquid crystal. This phase, which we call {\it compactified supersolid} (CSS),  originates from wrapping the basal planes of the bulk {\it hcp} solid into concentric cylindrical shells, with several central shells exhibiting superfluidity along the axial direction. Its main feature is the presence of a topological defect which can be viewed  as  Frank's disclination with index $n=1$ observed in liquid crystals, and which, in addition,  has a superfluid core. The  CSS as well as its transition to an {\it  insulating compactified solid} with a very wide hysteresis loop are found by {\it ab initio} Monte Carlo simulations. A simple analytical model captures qualitatively correctly the main property of the CSS -- a gradual decrease of the superfluid response with increasing pressure.  
           
\end{abstract}

\maketitle
The emergence of unexpected phenomena in simple systems is one of the central themes in physics. A  historic example is  \he4 consisting of structureless bosons 
which, in addition to the classical phases, can exhibit macroscopic quantum behavior -- superfluidity. Whether the crystalline and superfluid orders can occur simultaneously and form a  supersolid is a question that still captivates the community  45 years after it was proposed \cite{ssolid}. While no supersolid has been seen in ideal {\it hcp} samples, some grain boundaries and dislocations have been found to support superfluidity in {\it ab initio} simulations \cite{grain,screw} and, possibly, in the experiment \cite{Hallock}. A metastable phase, superglass, has also been observed in the simulations \cite{superglass}.    

In apparently different fields, the emergence of quantum liquid crystals~\cite{Kivelson98} has been proposed in such contexts~\cite{Fradkin} as the quantum Hall effect, bilayer Sr$_3$Ru$_2$O$_7$,  the cuprates, and  highly magnetic  dipolar degenerate fermionic cold atoms~\cite{Fregoso2009} such as Cr~\cite{Griesmaier2005}, Er~\cite{McClelland2006}, and Dy~\cite{Lu2012}, and for population imbalanced Fermi gases~\cite{Radzihovsky}. The role of curved substrate in inducing novel 2D phases was discussed in Ref.\cite{Nelson}. 

In this Letter we reveal  a phase induced by geometrical confinement, the {\it compactified supersolid}.
This phase features topological properties of a  liquid crystal as well as the quantum phenomenon of superfluidity.  Our {\it ab initio} simulations show that CSS must occur in \he4 confined to materials with a cylindrical geometry with mesoscopic diameter as large as 30nm (see below), that is, in vycor glass or in artificially made nanopores.  Due to its topological nature, the CSS is robust against smooth deformations of the pores or disorder, implying that simulations inside an ideal cylinder are sufficient for elucidating its main features. There are also, as we will see, experimental signatures consistent with its existence. The description of CSS as well as of the  {\it compactified solid} (CS) naturally invokes variables, objects and terminology typical for  liquid crystals. These are the smectic-A type {\it layers} with the local hcp axis playing the role of the nematic-type {\it director} characterized by the {\it splay} and forming Frank {\it disclination} with index 1 (see, e.g., Ref.~\cite{LandauLifshitz}).

We start with discussing the similarity between the roton-induced spatial density modulation in superfluid \he4 close to a hard wall \cite{roton} and layers in classical smectic-A liquid crystals.  Such a modulation as well as the liquid crystal layers both exhibit zero shear response in the tangential directions. If the hard wall has cylindrical shape, the crests and troughs of the  modulation  acquire the cylindrical shape and form a structure containing the Frank disclination  observed in liquid crystals (see   Fig.2a  in Ref.~\cite{deGennes}). At high pressure, the modulation transforms into shells of the CS hereby freezing the disclination with its long-range splay. This splay may partially melt a few shells in the vicinity of the disclination line resulting in the CSS. This mechanism is similar to the strain-induced superfluid core of some dislocations in {\it hcp} \he4  \cite{stress}.   There is, though, a significant difference between the two:  In contrast to dislocations,  the disclination is a part of the {\it ground state} of the CSS and CS. In our simulations we have observed both phases as well as the transformation between them characterized by a very wide hysteresis which implies that rather long-lived metastable superfluidity can exist at  pressures much higher than in macroscopic 3D samples of solid \he4.

The compactified structural order of \he4 has previously been observed numerically. A variational study \cite{Reatto} has found that \he4 forms shells concentric with the pore wall. These shells are hexagonal layers rolled into cylinders which are claimed to be always superfluid.
{\it Ab initio} Monte Carlo (MC) simulations~\cite{DelMaestro11} at saturated vapor pressure have equally found the shell structure, but with no intra-shell structural order.  While a pore with a diameter  $R_0=2.9$\AA~  is insulating, a pore with $R_0=14$\AA~ demonstrates weak superfluidity.

{\it Model description}. Here we introduce the relevant coarse grained variables and sketch the description of the main features of CSS and CS. 
The key variable is the envelop $\vec{C}(\vec{r})$ of the gradient of the density modulation at the roton wavevector $k_r$. In the liquid phase \he4 is characterized by a structure factor with the peak at $k_r$. In real space such a peak implies that the boundary induces spatially decaying density oscillations $ \rho'(\vec{r}) \sim \exp(- i \vec{k} \vec{r}) + c.c. $ with $|\vec{k}| \approx k_r$   and the exponentially decaying part ($k$ has an imaginary part) determined by the roton gap \cite{roton}. In a cylindrical geometry the modulation picks up the cylindrical symmetry $   \rho'(r) \sim \exp(i k_r r)+c.c. $, where $r$ is the radial coordinate.  Accordingly,  $\vec{C}(\vec{r})\sim k_r \vec{r}/r$ winds around the cylinder axis in the same manner as the {\it director} field does in a  liquid smectic-A crystal containing Frank disclination with the index $n=1$. At high pressure, the modulations become crystalline shells which, in addition to the director field $\vec{C}$ setting the local orientation of the hcp axis, must be also described by the intra-shell (quasi-) hexagonal order.

Similarly to liquid crystals, the contributions of $\vec{C}$ to the energy can be chacterized by  {\it splay},  {\it twist} and {\it bend} as well as by shell deformations.  Given the simplest geometry,  we ignore {\it twist} and {\it bend} and consider only the {\it splay} energy  
\beq
E_s \sim \int d^2r dz (\vec{\nabla} \vec{C})^2 \sim \ln (R_0/R) L_z.
\label{Es}
\eeq
Here, $z$ is the coordinate along the cylindrical axis, $L_z$  stands for the total length of the cylinder, $R_0$ denotes the cylinder radius, and $R<R_0$ is the radius of the disclination core inside which the splay singularity $ \vec{\nabla} \vec{C} \sim 1/r$ has been  cured by melting the inner shells into a superfluid (characterized by a complex field $\psi$ as another order parameter). 
Thus, while being of the order of the interparticle distance in the CS phase with $\psi=0$,  $R$ can be mesoscopically large in the CSS  phase so that there is $\psi \neq 0$ inside the core. In the simulations we associate the CSS to CS transition with the vanishing of superfluidity. 

Melting of the core above the melting pressure  $P_m \sim 25$ bar costs energy
$E_c\approx (\mu_l -\mu_s) R^2 L_z$, where $\mu_l > \mu_s $ stand for the chemical potentials of  liquid and solid, respectively. Thus, the equilibrium solution for the core radius can be found by minimizing the total CSS energy $E_s +E_c$ with respect to $R$. This gives
$ R\propto 1/\sqrt{\mu_l -\mu_s}$ where we ignore the surface tension and assume the limit $R \ll R_0$. 
As the external pressure $P$ increases above $P_m$, the core radius decreases and so does the superfluid response  $\rho_s \propto R^2$:
\beq
\rho_s \propto   \frac{1}{P-P_m} , 
\label{RR}
\eeq  
where we used $\mu_l -\mu_s \sim P-P_m$. Eq.~(\ref{RR}) is qualitatively consistent with the simulations (see Fig.~\ref{fig:energy}) and with the experimental observations \cite{Reppy} of \he4 in vycor.

{\it Ab initio simulations --}
We have conducted {\it ab initio}  MC simulations ( by the worm algorithm ~\cite{worm1, worm2}) of  a grand-canonical ensemble at various values of chemical potential $\mu$ so that there are $N \sim 600- 2000$ $^{4}$He atoms  confined inside a cylindrical volume with periodic boundary conditions along the $z$-direction ($L_z=30$\AA) at  temperature $T=0.2$K. In the Hamiltonian, 
\begin{equation}
H = - \frac{\hbar^2}{2m} \sum_{i=1}^{N} \vec{\nabla}^2_i + \sum_{i<j}V_{\rm Aziz}(r_{ij}) + \sum_{i} V_{\rm sub}(\vec{r}_{i}),
\label{eq:hamiltonian}
\end{equation}
$\frac{\hbar^2}{2m}\vec{\nabla}^2_i$  is the kinetic energy operator of $i$-th $^4$He atom located at $\vec{r}_i$; $V_{\rm Aziz}(r_{ij}) $ is the standard central Aziz-potential~\cite{Aziz}, with  $ r_{ij}\equiv |\vec{r}_i - \vec{r}_j|$. The potential $V_{\rm sub}=  \frac{D}{2} \left(\frac{b^9}{\xi^9}  - 3 \frac{b^3}{\xi^3}  \right)$, with  $ b=2.0$\AA and $ D=80$K, acts  between the 
pore wall and \he4 atoms. It is the so called 3-9 potential \cite{DelMaestro11}, where in the cylindrical geometry $\xi = R_0 - r>0$, with $R_0=25.8$\AA. The precize shape of $V_{\rm sub}(\bm{r}_{i})$ does not change anything qualitatively  (cf. Ref.~\cite{Boninsegni10})  as long as its depth $D=80$K is much bigger than that ($\approx 11$K) of $V_{\rm Aziz}(r_{ij}) $.

\begin{figure}[t]
\centerline{\includegraphics[width=1.2\columnwidth, angle=0]{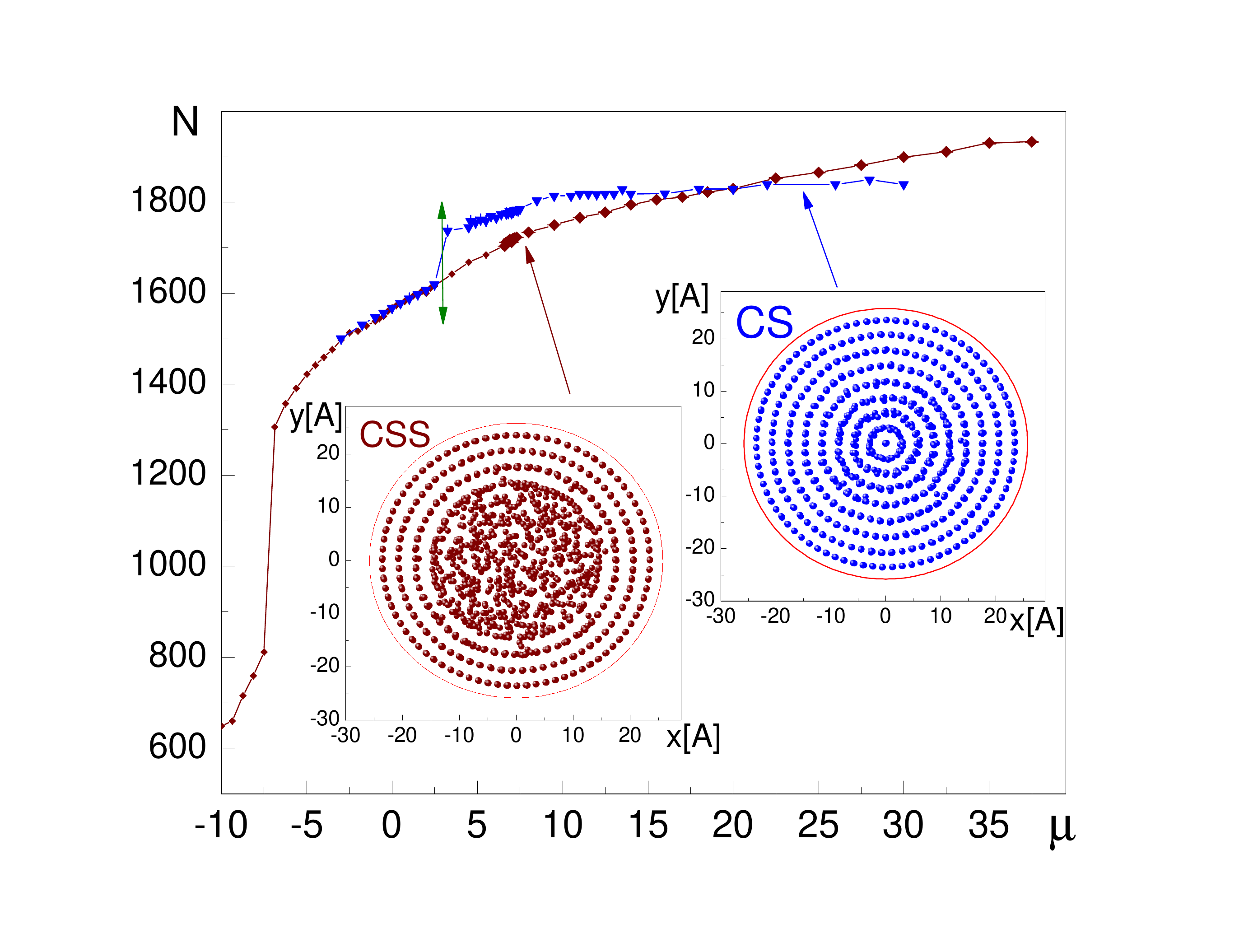}}
\caption{(Color online) The number of particles $N$ vs chemical potential $\mu$ for two phases CSS and CS. The double-sided arrow indicates the ending of the hysteresis at $\mu\approx 3.2K$. Insets: columnar view along the cylindrical axis of a  typical atomic configuration of CSS (left) and CS (right) both at $\mu=7.1$K.  The solid (red) circle outlines the pore boundary at $R=R_0$. The quasi-disordered region in the CSS is the superfluid core. }
\label{fig:nmu}
\end{figure} 
\begin{figure}[t]
\centerline{\includegraphics[width=1.2\columnwidth, angle=0]{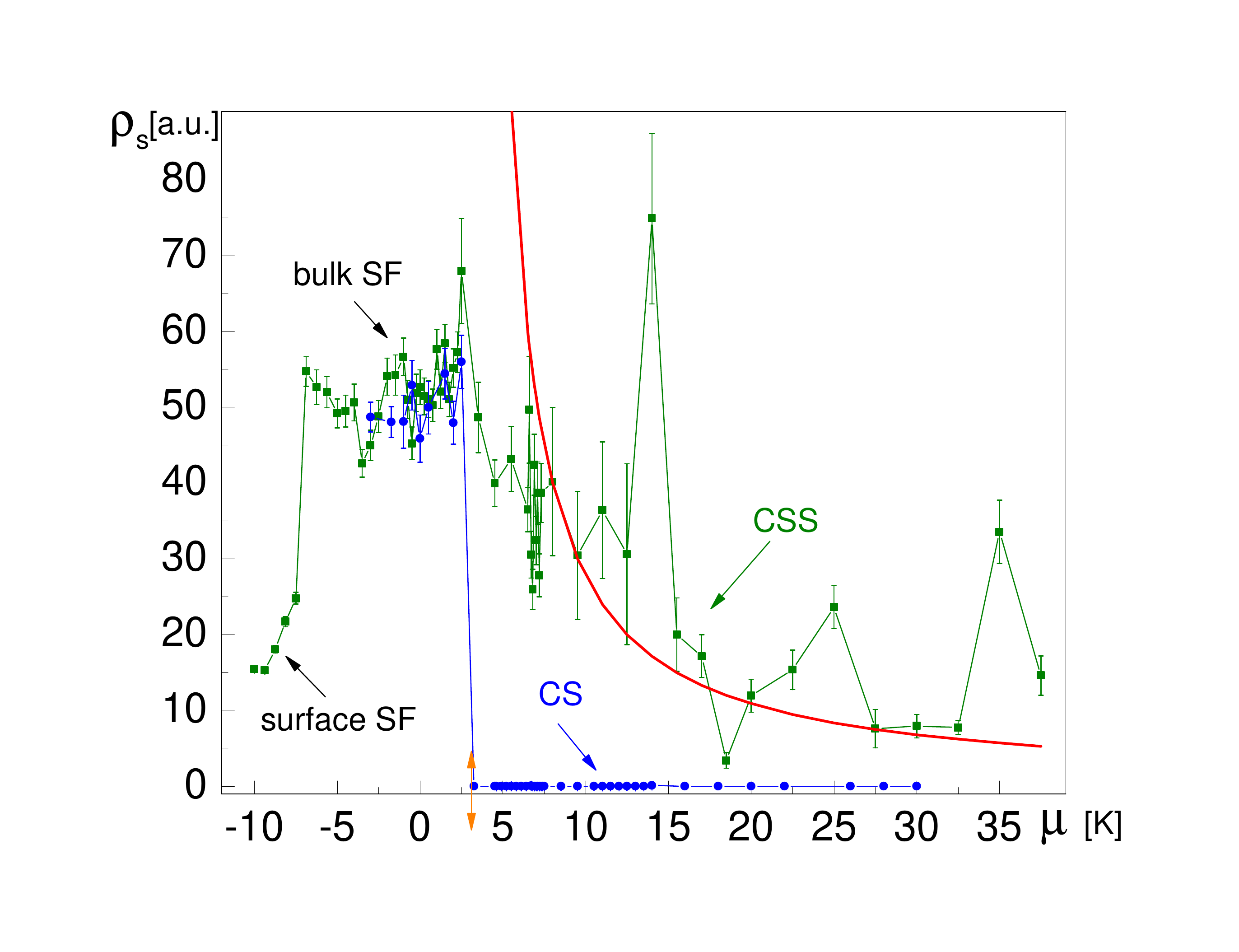}}
\caption{(Color online)  Superfluid stiffness $\rho_s$ vs $\mu$  of \he4 in the nanopore. 
The solid red line is the fit by Eq.~(\ref{RR}). 
The double-sided arrow indicates the closing of the hysteresis loop.
}
\label{fig:energy}
\end{figure}
\begin{figure}[t]
\centerline{\includegraphics[width=1.2\columnwidth, angle=0]{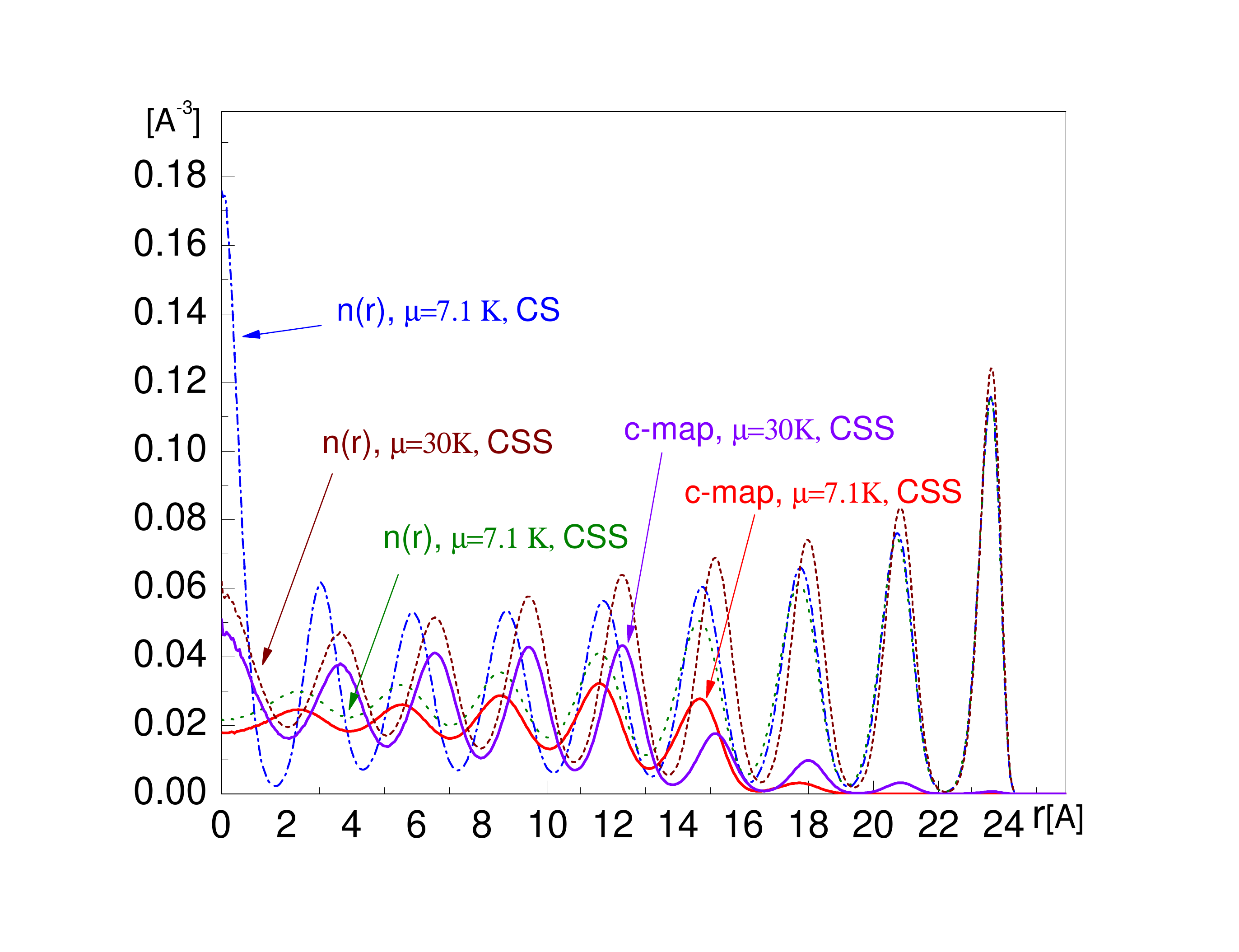}}
\caption{(Color online) Atomic, $n(r)$, and condensate, c-map, densities along the radial direction in the samples corresponding to $\mu=7.1$ and $30$K.   }
\label{fig:den_win}
\end{figure}

A typical atomic configuration of the CSS, shown in the left inset in Fig.~\ref{fig:nmu}, features well defined outer shells (3 of them at $\mu=7.1$K), each with slightly distorted hexagonal order (shown in Figs. 4,5  in the Suppl. Mat.~\cite{suppl}), as well as the superfluid core which is visibly disordered (within the radius $R \sim 12-15$\AA). 
Most of the superfluid response seen in Fig.~\ref{fig:energy} comes from this core. Despite being apparently fully disordered, there are distinct radial density $n(r)$ as well as superfluid density (represented by the so called condensate map or c-map, see  in Ref.\cite{superglass}) modulations in the core seen in Fig.~\ref{fig:den_win}.   
Increasing $\mu$ in the CSS phase leads to
the compression of the superfluid core and to the gradual suppression of the superfluid stiffness $\rho_s$.
The core compression can be recognized in Fig.~\ref{fig:den_win}: The concentration of the c-map in the center is higher in the $\mu=30$K sample than in the $\mu=7.1$K one.
The red line in Fig.~\ref{fig:energy} is the fit by Eq.~(\ref{RR}) of the numerically found $\rho_s$, where we have used
$(P-P_m) \propto  (\mu -\mu_m)$, with $\mu_m$ corresponding to the melting of macroscopic {\it hcp} samples (with no disclination). 
In order to find $\mu_m$, we ran simulations in the slab geometry, that is, with a flat smooth wall and periodic boundary conditions along the wall producing the same 3-9 potential $V_{\rm sub}$. We  found the solid spinodal at $\mu=\mu_{sp} \approx 3.0$K, and the liquid spinodal at about 7K.
In the experiment \cite{Souris2011} it has been determined that the solid spinodal pressure is below $P_m$ by about 10-15\%.  Thus, we estimate $\mu_m \approx 1.15\mu_{sp}\approx 3.5$K. 
On top of the overall suppression of $\rho_s$ vs $\mu$ in Fig.~\ref{fig:energy} consistent with Eq.~(\ref{RR})  there are additional peaks and dips in $\rho_s$ vs $\mu$ (see Fig.~\ref{fig:energy}), which may be related to structural fluctuations caused by the proximity to the CS phase.

The emergence of various phases in the pore is reflected in the dependence of the particle number $N$ vs $\mu$ in Fig.~\ref{fig:nmu}.
At $\mu < - 50$K,  the outermost shell becomes populated and forms a superfluid.
 It solidifies into a hexagonal (insulating) shell at $\mu \approx -30$K. [ This stage is not reflected in Fig.~\ref{fig:nmu}]. The second shell forms in the range $ -12< \mu < -7$K. It is a low density surface superfluid (SF) which exhibits no  visible structural order (see Fig.1 of the Suppl. Mat. \cite{suppl}). Accordingly, the curves in Figs.~\ref{fig:nmu},\ref{fig:energy}  show linear dependencies on $\mu$ in this range. During this stage the pore bulk remains empty.  At $\mu \approx -7$K \he4 undergoes a dimensional crossover  marked by the jumps in $N$ (Fig.~\ref{fig:nmu}) and in the superfluid stiffness $\rho_s$ ( Fig.~\ref{fig:energy}): at $\mu > -7$K the whole pore becomes filled by \he4 forming a low density superfluid.  In this phase, while only two outer shells are clearly defined and possess hexagonal order, the weak radial density modulations induced by the roton \cite{roton} can also be detected in the pore bulk (see Fig.2 in the Suppl. Mat.~\cite{suppl}).

 The CS begins as a metastable phase at $\mu \approx 3.2$K as shown in Fig.~\ref{fig:nmu}. The shells (we observed eigth of them) of the CS are well-defined and exhibit hexagonal order consistent with the whole {\it hcp} crystal being compactified (see the Suppl. Mat.~\cite{suppl} for details). There is also a central (insulating) core hosting \he4 atoms along a very narrow straight line coinciding with the cylinder axis.  The CS phase is characterized by zero superfluid response $\rho_s=0$ as seen in Fig.~\ref{fig:energy}. A weak dependence of $N$ vs $\mu$ of the CS shown in Fig.~\ref{fig:nmu}  indicates that doping is still possible in this insulating state. However, the extra particles (or vacancies) do not form a superfluid. Instead, they phase separate,  very similarly to the case of macroscopic samples studied in Ref.~\cite{fate}. Lowering $\mu$ below $\mu \approx 3.2$K results in a jump-like melting of the CS into the bulk SF (which gradually transforms into the CSS as $\mu$ increases). This indicates closing of the hysteresis at its low end as marked  by the double sided arrows  in the curves  $N$ vs $\mu$ (Fig.~\ref{fig:nmu}) and in  $\rho_s$ vs $\mu$ (Fig.~\ref{fig:energy}). 

While the CS is metastable at $ 3.2$K$<\mu < 7-10$K, the CSS is stable in this region and becomes metastable above $\mu \approx 7-10$K. Due to the very wide hysteresis a more accurate finding of the transition point turned out to be very challenging. As Fig.~\ref{fig:nmu} indicates, the upper end of the hysteresis, where the metastable CSS transforms into the stable CS, could not be determined: the CSS persisted at $\mu$ as high as 38K, in sharp contrast to the results in the slab geometry with the hysteresis loop being only 4K wide (see above).

{\it Discussion}.
One of the longstanding open questions is the nature of solid \he4 in a vycor. 
Superfluidity there persists at a pressure $P$ as high as 10-20 bar above the melting pressure. Several models have been proposed to explain this effect \cite{Dash,Beamish1983,Reppy,Adams}, including the conjecture that \he4 remains liquid close to the vycor wall with the solid forming away from the wall \cite{Reppy}. MC simulations of about 200 \he4 atoms \cite{Khairallah05} with the artificially fixed hcp solid at some small distance from the wall support this picture. As our simulations of bigger samples in a realistic geometry show, there is no liquid layer adjacent to the wall, and, instead, there is a liquid core at the pore center. We also note that our observations are in contrast to the variational results \cite{Reatto} predicting that the solid in a nanopore is always a supersolid.

The wetting models \cite{Dash,Beamish1983,Reppy}, where \he4 at the wall remains liquid until pressure overcomes the surface tension nucleation barrier, encounter troubles explaining the gradual decrease of the superfluid response with pressure \cite{Reppy} because the nucleation mechanism implies an abrupt solidification. In contrast, the CSS is characterized by a gradual decrease of its superfluid response with pressure. The experimental observation  of the overall decrease  of entropy of the liquid part of \he4 in vycor with increasing pressure, seen in Fig. 1c of Ref.~\cite{Yamamoto},  is also consistent with the shrinking of the superfluid core with pressure observed in our simulations.

Our analysis and simulations of the topological phases of \he4 in nanopores are directly relevant to pores with radii below a threshold, $R_{\rm max} \sim 300$\AA~ (as estimated in  the Suppl. Mat.~\cite{suppl}), well above typical radii in vycor or gelsil glass. We consider it  a lower bound because the CS or CSS may exist as metastable phases in much larger pores due to the geometrical (macroscopic) energy barrier between the compactified and standard $hcp$ solids. This implies that the CSS can be grown and studied in a more controlled way in artificially created pores. 

\begin{figure}[t]
\centerline{\includegraphics[width=1.2\columnwidth, angle=0]{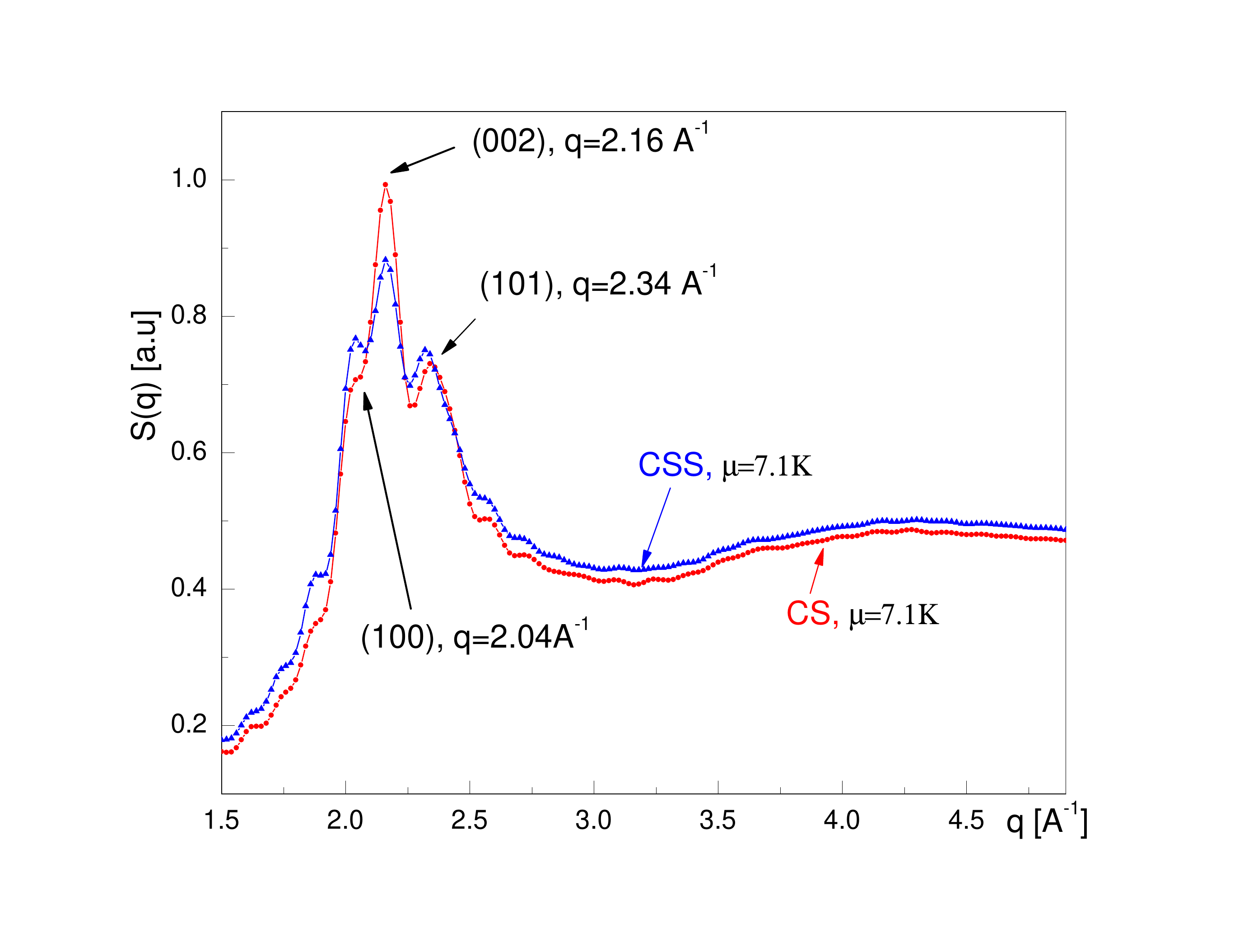}}
\vspace*{-0.5cm}
\caption{(Color online) Structure factors of CSS and CS averaged over orientations. The major peaks are labelled according to the standard $hcp$ classification.}
\label{fig:structure}
\end{figure}
In the recent experiment \cite{Mulders2013}, \he4 in vycor was found to be in the {\it bcc} phase at $P<98$ bar and $T \approx 0.5-0.7$K, whereas the transformation to the $hcp$ solid takes place at higher pressure. 
The structure factor for CS and CSS (averaged over all orientations) found in our simulations and shown in Fig.~\ref{fig:structure} is strikingly similar to the one found in Ref.\cite{Mulders2013} at high pressure.
 It features three main peaks in the momentum region  $\sim 2.0-2.2$\AA$^{-1}$: one strong and two satellite peaks reminiscent of the three main Bragg peaks of {\it hcp} solid. The higher order peaks are washed out by quantum fluctuations and are "hidden" under the wide shoulder at high momenta. 
  In future work it would be important  to repeat the experiment \cite{Mulders2013} at lower temperatures, as well as to perform the MC simulations at temperatures higher than $T=0.2$K. One possibility is that there is a non-trivial transition line in the $P-T$ plane where the compactified {\it hcp} solid becomes a compactified {\it bcc} solid. 


Finally, we suggest  (and leave the analysis for future work)  that the CSS disclinations ending at the interface of vycor and the bulk solid \he4 may attract (and also create) the bulk dislocations with superfluid cores~\cite{screw}, so that the superflow through the bulk becomes possible as observed in Ref.~\cite{Hallock}.

{\it Conclusion} -- 
When \he4 is subjected to geometrical confinement with cylindrical topology, it can be found in the compactified solid and compactified supersolid phases. Both are characterized by the shelled structure reminiscent of smectic-A liquid crystal containing Frank's disclination. While the CS is insulating, the CSS exhibits superfluid response within the melted core of the disclination. Such a core  can persist in a metastable state at pressures significantly exceeding the spinodal for the overpressured superfluid in macroscopic samples of \he4. 
This finding offers a compelling explanation for the physics of \he4 confined to restricted geometries at high pressure where the local $C_6$ axis, playing the role of the nematic director, can not  be uniquely defined everywhere.  Thus, in the multiple-connected geometry of nanoporous materials confining \he4 the superfluid response at high pressure should be controlled by a network of the disclinations.

Acknowledgements -- We wish to thank  M. Boninsegni, A. Del Maestro,  R. Hallock, N. Mulders, and Boris Svistunov for fruitful discussions. This work was supported by FP7/Marie-Curie Grant No. 321918 (``FDIAGMC"),  FP7/ERC Starting Grant No. 306897;  by the NSF grant PHY1314469, and by the grant from CUNY HPCC  under NSF Grants CNS-0855217, CNS-0958379 and ACI-1126113.

\vspace*{3cm}

{\huge{Supplemental Material}} \\

In the Supplemental Material we illustrate graphically  typical atomic configurations, the density profile of the surface layer phase and of the pore filled with superfluid. 
We also look in more detail into the compactification process by computing the interparticle distances, the strain field and comparing the energies of the compactified and non-compactified structures in order to establish the stability condition.

\section{The surface and the low density phases}
\begin{figure}[h]
\centerline{\includegraphics[width=1.2\columnwidth, angle=-0]{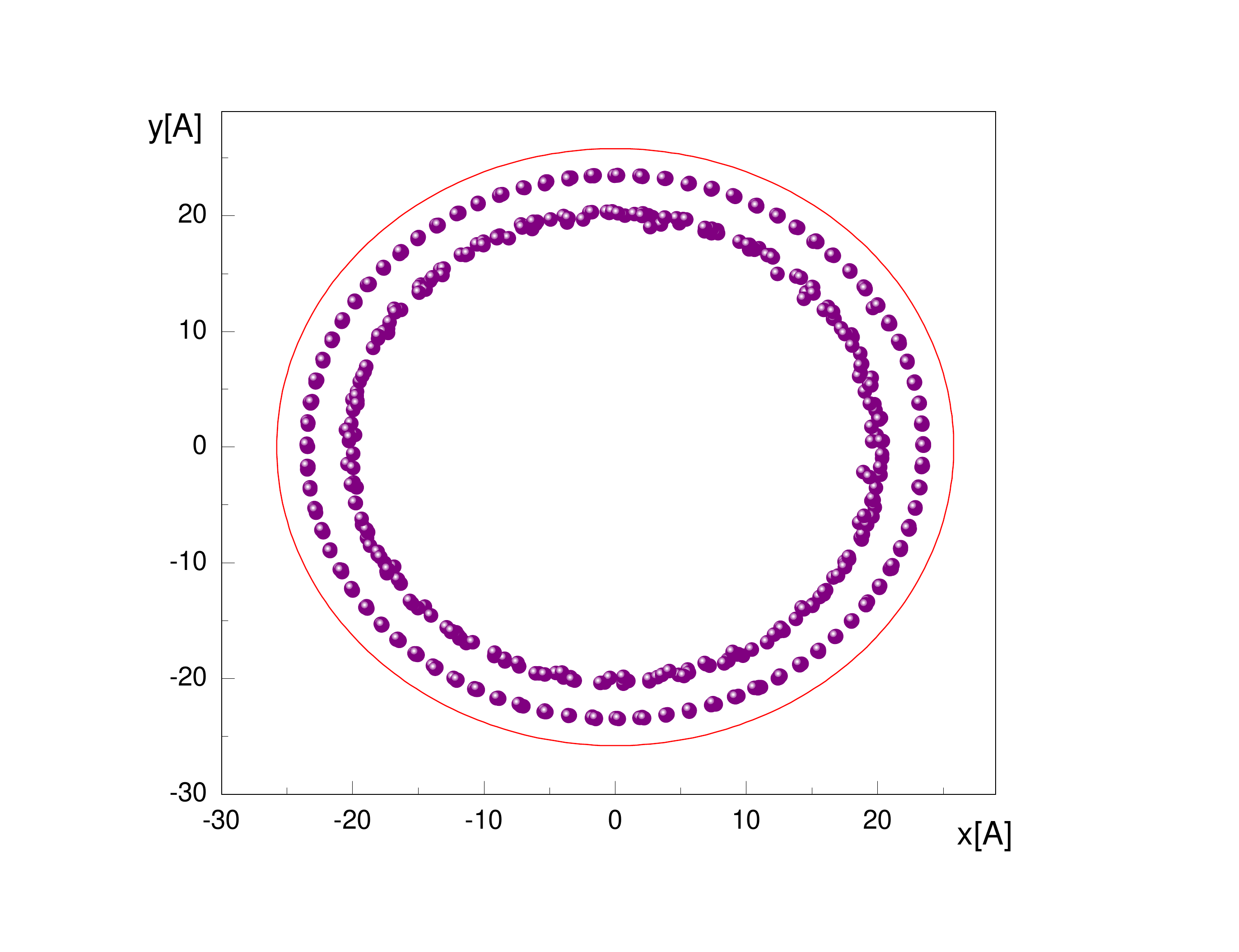}}
\vspace*{-0.5cm}
\caption{(Color online) The columnar view  along the pore axis of a typical atomic positions in the $\mu=-9.375$K sample.  The red circle marks the position of the hard wall.}
\label{fig:2shells}
\end{figure}

At low $\mu$, at most, the first two outer shells are formed. The columnar view of such a typical atomic configuration (at $\mu =-9.375$K) is shown in Fig.~\ref{fig:2shells}. In this phase the superfluid response comes from the second shell (farthest from the wall). Accordingly, the first shell is ordered and the second one is disordered.
\begin{figure}[h]
\centerline{\includegraphics[width=1.2\columnwidth, angle=-0]{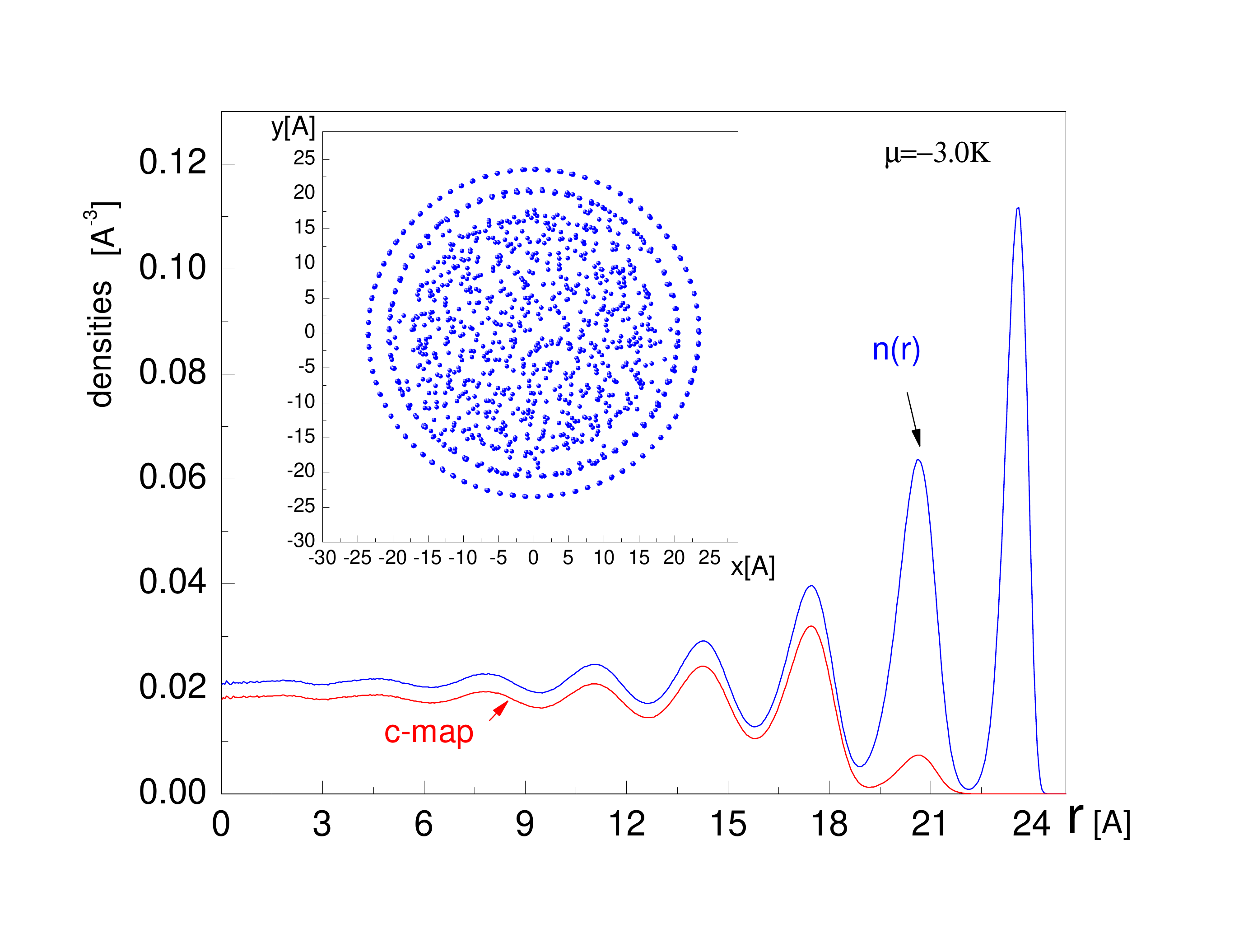}}
\vspace*{-0.5cm}
\caption{(Color online) The density modulations $n(r)$ and the c-map in the sample, $\mu=-3.0$K, where the pore bulk is occupied by a low density superfluid. The first two strong peaks in $n(r)$ correspond to the two ordered surface shells, and the weaker peaks are induced by the roton feature of the spectrum.   Inset: the columnar view of a typical configuration along the pore axis, $\mu =-3.0$K.  The radial density modulations in the bulk, $ r< 18$\AA, cannot be distinguished visually.}
\label{fig:roton}
\end{figure}  
The bulk phase exists at $\mu \geq -7$K. It can be viewed as two outer shells coexisting with the low density superfluid filling the pore bulk. The bulk density $n(r)$ and the superfluid density (shown by the c-map) are both modulated in the radial direction at the wavelength corresponding to the roton. These modulations observed in a sample $\mu=-3.0$K are shown in Fig.~\ref{fig:roton}. These are the precursors of the shells which eventually form the CS and CSS. 

\section{The CS vs non-CS geometries}
\begin{figure}[h]
\centerline{\includegraphics[width=1.5\columnwidth, angle=-0]{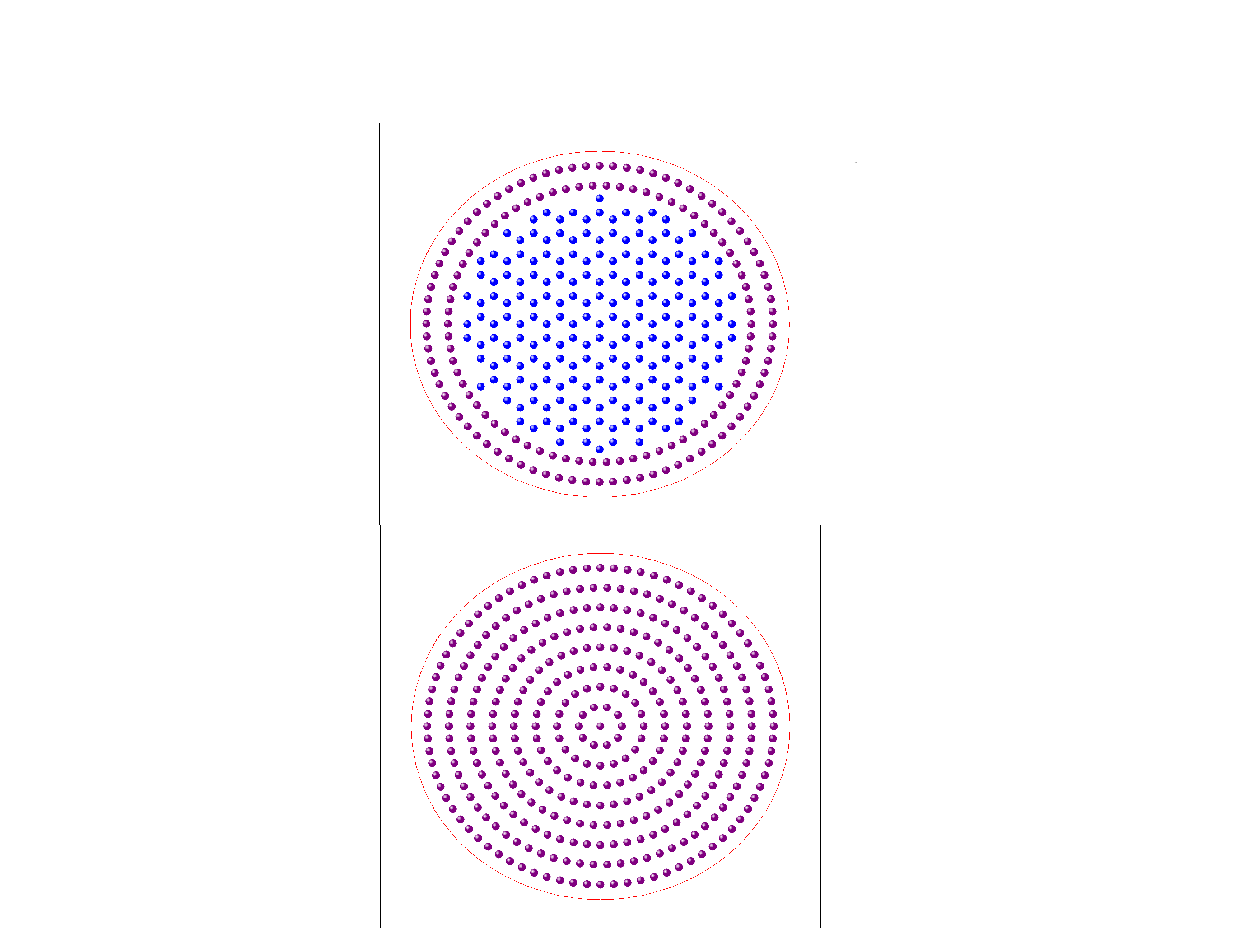}}
\caption{(Color online) The columnar views  along the pore axis of two possible configurations. Top panel: the non-compactified $hcp$ solid. The $C_6$ axis is in the radial direction in the outer shells (purple dots) and it becomes along the cylinder axis (perpendicular to the page plane) in the inner part  of the pore (blue dots). Bottom panel: The compactified $hcp$ solid. The $C_6$ axis is along the radial direction (in the page plane) in the whole sample. It winds in a manner similar to the director in the Frank  nematic disclination with the index $n=1$ (see in Ref. \cite{LandauLifshitz}).}
\label{fig:CS_S}
\end{figure}

A possible fitting of the $hcp$ structure into a cylinder with least of the bulk strain  is shown in the top panel of Fig.~\ref{fig:CS_S}. In this structure, the strong attractive wall potential creates several (here we show two) outmost hexagonal shells wrapped around the wall. Since being closely packed in 2D, such shells minimize the surface energy. The $C_6$ axis in these shells is oriented radially with respect to the pore axis. In the inner part of the pore, however, the $C_6$ axis is aligned with the cylinder axis similarly to the director in the non-singular nematic disclination solution (see in Ref.~\cite{LandauLifshitz}). 
Simulations of pores with radii $< 15$\AA \cite{Reatto,DelMaestro11}  as well as our present work with the pore of almost twice that radius show that  this configuration is not realized at least in pores with radii less than $\sim$ 30\AA. The preferred configuration is the compactified $hcp$ solid shown in the bottom panel in Fig.~\ref{fig:CS_S}. This configuration hosts the Frank disclination of index $n=1$, with its core coinciding with the cylinder axis at $r=0$. There is a string of  atoms arranged along a very narrow line at $r=0$. 
  
As we will estimate below, the compactified configuration, CS, has lower energy than the standard $hcp$ in pores with radii, at least, up to $R_0 \sim 300$\AA.
\begin{figure}[t]
\centerline{\includegraphics[width=1.1\columnwidth, angle=-0]{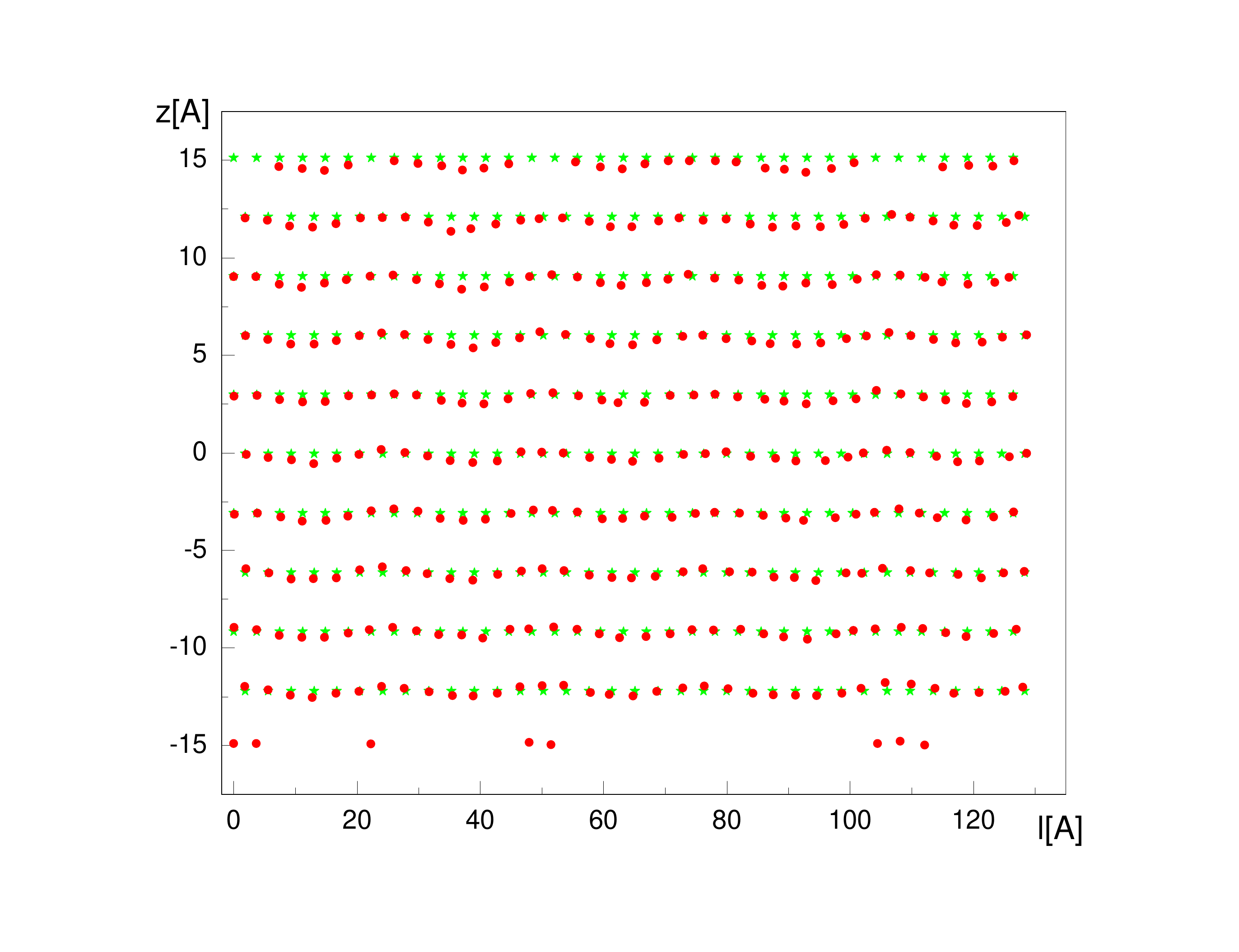}}
\vspace*{-0.5cm}
\caption{(Color online) The unrolled second outer shell (counted from the wall). The green stars show the ideal CS positions and the red dots are the atomic positions from a typical configurations from the simulations. The vertical axis is the cylindrical z-coordinate, and the horizontal axis $l$ is the coordinate along the shell circumference. The five-fold angular modulation along the z-axis with the amplitude $\approx 0.5$\AA ~ is clearly seen.}
\label{fig:unrolled}
\end{figure} 
\begin{figure}[h]
\centerline{\includegraphics[width=1.2\columnwidth, angle=-0]{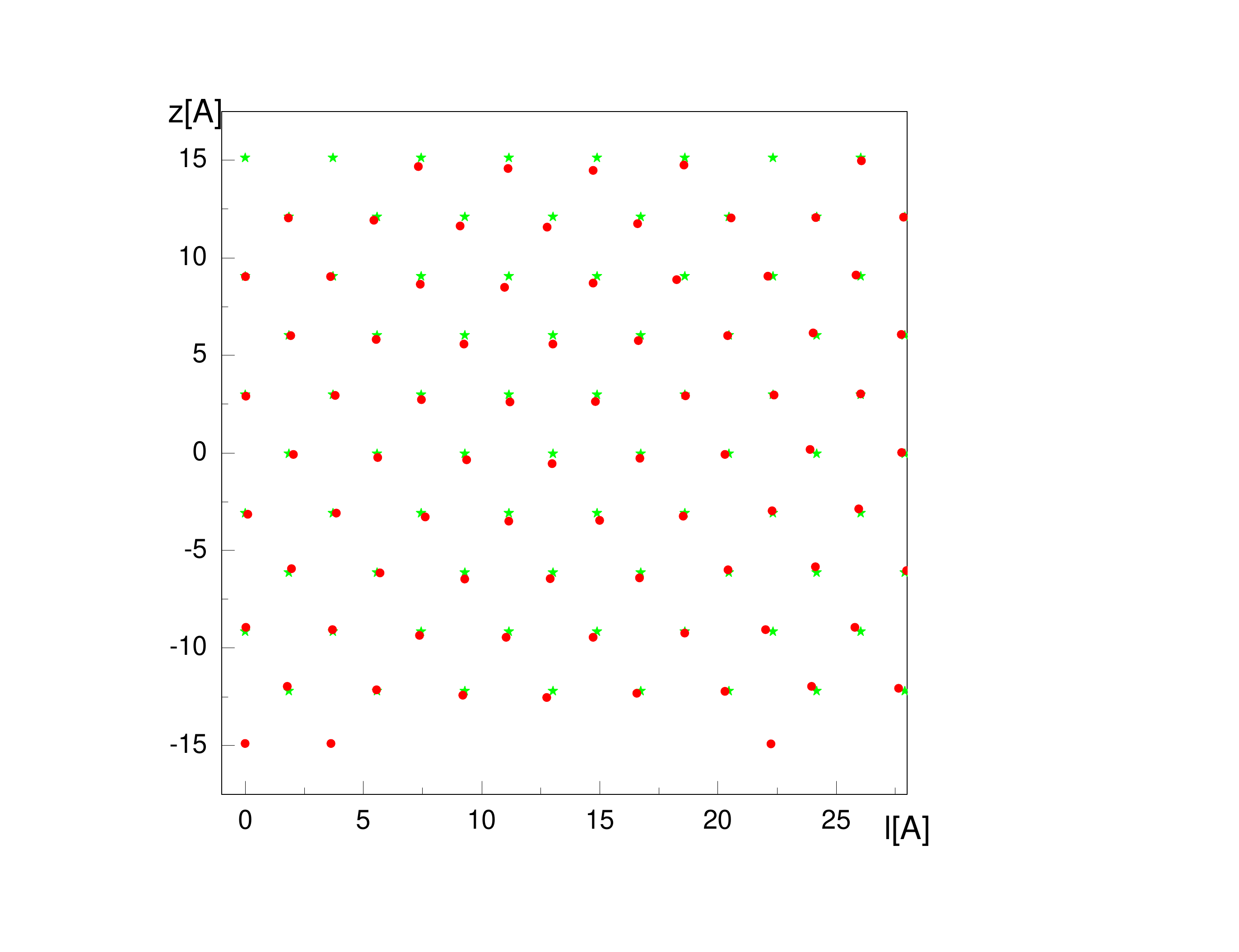}}
\vspace*{-0.5cm}
\caption{(Color online) The same pattern as in Fig.~\ref{fig:unrolled} is shown on smaller scale so that the triangular layer structure is obvious.}
\label{fig:unrolled2}
\end{figure} 

Let's consider in detail the ideal compactified $hcp$ geometry. In order to produce minimal residual strain the A-B hexagonal (basal) planes of the standard $hcp$ structure should be rolled into concentric cylinders along the direction of the elementary cell vector belonging to the basal plane so that the orthogonal direction is aligned with the cylinder axis. Eight such shells are seen in the bottom panel in Fig.~\ref{fig:CS_S} (plus the central core). The actual structure of the CS  found in the simulations is very close to the one formed by this procedure, as is demonstrated in Figs.~\ref{fig:unrolled} and \ref{fig:unrolled2} . 

The number of unit cells in the $N$th shell with radius $R_N$ is given by $M(N)= 2\pi R_N/a(N)$, where the length of the unit cell $a(N)$ may vary from shell to shell. The next shell  has radius $R_{N+1}=R_N + a_z(N)$, where $a_z(N)$ is the radial distance between the $N$th and $(N+1)$th shells. In a perfect {\it hcp} crystal $a_z = \sqrt{2/3} a$, where $a\approx 3.6-3.7$\AA ~ is the unit cell length in the basal plane. In the compactified {\it hcp} solid this relation needs to be relaxed to  $a_z(N)=\gamma(N) \sqrt{2/3} a(N)$ with $\gamma(N) \approx 1$ in order to minimize the strain. Thus, the radius of the $N$-th shell becomes $R_N= \sum_{N'=1}^N \gamma(N') \sqrt{2/3} a(N')$. Expressing $a(N)$ in terms of the integer number $M(N)$, we obtain the equation for the shell radii
\begin{equation}
R_N = 2\pi \sqrt{\frac{2}{3}} \sum_{N'=1}^N \gamma(N) \frac{R_{N'}}{M(N')}.
\label{shells}    
\end{equation}
This equation has a solution $M(N)=5N,\, \gamma(N)= 5/( 2\pi \sqrt{\frac{2}{3}})\approx 0.975$. Thus, $a_z(N)$ is compressed (radially)  by about $2.5\%$ when compared with the standard {\it hcp} crystal. 

In addition to the radial compression, there is shear strain of one shell with respect to its neighbor. The "quantization" rule $M(N)=5N$, Eq.(\ref{shells}) implies that the circumference of each shell is broken into 5 equal angular segments, each subtended by an angle $2\pi/5$. Within each segment, the smallest distance between two atoms from neigboring shells reaches the minimum $\sqrt{3}a/2\approx 0.87a$ along one radial line. [There are five of such lines forming the $C_5$ symmetric pattern]. Thus, this strain can be estimated as $\approx 1-0.87=0.13$ at its maximum, and about $0.13/2\sim 0.065$ on average for the whole sample.
In simulations we have observed that such strain has been relaxed to about $0.04$  by the static angular modulation of atomic displacement about $0.5$\AA~ along the pore axis with the angular period $2\pi/5$. Fig.~\ref{fig:unrolled} shows this pattern (see also Fig.~\ref{fig:unrolled2}).

Thus, the CS structure can be characterized by the $0.025$ compression strain and by about $0.04$ of the shear strain. 
We estimate the resulting energy change as being due to the elastic energy $\delta E_{el} \sim (0.025^2+ 0.04^4) E_D$, where $E_D$ is determined by elastic constants  defining the Debye energy $\sim 30$K of solid \he4.Thus, the compactification 
costs about extra $0.07$K per particle. In other words,  the non-compactified $hcp$ \he4 of the same average density represented in the upper panel in Fig.~\ref{fig:CS_S} has less energy $\sim 0.07KR_0^2$  if one ignores the boundary.     
The boundary between the ideal $hcp$ and the outer shells are characterized by maximal possible misfit: the $C_6$ axis 
must rotate by $90$ degrees in order to be aligned with the cylinder axis. We estimate the energy of such misfit as being larger than $\sim 0.1E_D\sim 3$K. Thus, the total excess energy of the CS can be written as $ \sim - 3K\cdot 2\pi (R_0/a) + 0.07K\cdot \pi (R_0/a)^2$. It becomes larger than that of the non-compactified structure at radii larger than $R_0\sim 90 a$. For typical values of $a$ this estimate gives about $R_0\approx 300$\AA. 

\end{document}